\documentclass[12pt]{article}
\usepackage{amsmath}
\usepackage{cite}
\usepackage{graphicx}
\usepackage{subfigure}
\usepackage{times}
\usepackage{multicol}
\usepackage{fullpage}

\begin{document}
\pagestyle{empty}

\begin{center}
{\LARGE \bf Pseudoscalar Mass Spectrum in a Soft-Wall Model of AdS/QCD}

\vspace{1.0cm}

{Thomas M. Kelley}\footnote{E-mail: kelley@physics.umn.edu}, 
{Sean P. Bartz}\footnote{E-mail: bartz@physics.umn.edu},
{\small and}
{Joseph I. Kapusta}\footnote{E-mail: kapusta@physics.umn.edu}\\

\vspace{.5cm}
{\it\small {School of Physics and Astronomy, University of Minnesota\\
Minneapolis, Minnesota 55455, USA}}\\

\vspace{.3cm}
\today
\end{center}

\vspace{1cm}
\begin{abstract}
The Anti-de Sitter Space/Conformal Field Theory (AdS/CFT) correspondence may offer new and useful insights into the nonperturbative regime of strongly coupled gauge theories such as quantum chromodynamics (QCD). Recently a modified soft-wall AdS/QCD model incorporated independent sources for explicit and spontaneous chiral symmetry breaking and linear confinement. This model contains a modified dilaton and higher-order interaction terms in the Lagrangian. Within this model we explore the radial pseudoscalar mass spectrum using two different representations of the pion field. We find the mass eigenvalues associated with each representation, show the equivalence between the two, and find good agreement with the pion masses. The Gell-Mann--Oakes--Renner (GOR) relation is naturally obtained.
\end{abstract}

\vfill
\begin{flushleft}
\end{flushleft}
\eject
\pagestyle{empty}
\setcounter{page}{1}
\setcounter{footnote}{0}
\pagestyle{plain}

\section{Introduction}
\label{secIntro}

The theoretical framework for Anti-de Sitter Space/Conformal Field Theory (AdS/CFT) correspondence, relating type IIB string theory in $AdS_{5}\times S^{5}$ to $\mathcal{N}=4$ Super Yang-Mills (SYM) theory, was laid out in \cite{Maldacena:1997re, Gubser:1998bc, Witten:1998qj, Klebanov:1999tb}. This correspondence led to an effective dictionary relating strongly coupled gauge field theories to higher-dimensional, weakly coupled gravity theories. Such a correspondence removes the main difficulty of strongly coupled systems, the inapplicability of perturbation theory. Many papers have constructed dual models to capture part of the nonperturbative regime of gauge theories \cite{Erlich:2005qh, DaRold:2005zs, Karch:2006pv, Kwee:2007nq, Cherman:2008eh, Colangelo:2008us}. The majority of these use the correspondence dictionary to describe quantum chromodynamics (QCD)-like theories containing a large number of colors $N_{c}$.

Two methods are employed to formulate models using AdS/CFT correspondence: top-down and bottom-up. A top-down method uses a version of string theory to calculate an effective Lagrangian that, it is hoped, will contain certain key characteristics of QCD. A bottom-up approach, commonly known as AdS/QCD, uses the basic tenets of QCD in $d$ dimensions to formulate a dual gravity theory in AdS$_{d+1}$. Of course, constructing an AdS dual theory that encompasses the richness of QCD presents the greatest challenge, a task yet to be accomplished. In this paper, we implement a bottom-up approach evolving from the work of \cite{Erlich:2005qh, DaRold:2005zs, Karch:2006pv, Gherghetta:2009ac}, often associated with the term holographic QCD. This phenomenological model incorporates many of the crucial aspects of QCD. While the exact gravity dual is not known at present, nor even whether one exists, these types of models spark interest because of the potential of uncovering universal properties of strongly coupled gauge theories. 

We continue to study the model presented in \cite{Gherghetta:2009ac} by exploring the pseudoscalar sector. That model improves upon the previous ones both by incorporating confinement and separating explicit from spontaneous chiral symmetry breaking as in QCD.  We are particularly intrigued by this problem because the pseudoscalar representation used in \cite{DaRold:2005zs, DaRold:2005vr} {\it appears} to be incompatible with the representations used in \cite{Erlich:2005qh, Grigoryan:2008cc, Kwee:2007nq, DeFazio:2008mb}. Both are in the exponential form 
\begin{equation}
X = S {\rm e}^{i P},
\end{equation}
where $P$ is the pseudoscalar part of a complex field $X$. The former papers define $P$ using the vacuum expectation value $v(z)$ of the scalar field, 
$S(x,z)$, as $P = \pi^a(x,z)t^a/v(z)$, while the latter papers specify $P = 2\pi^{a}(x,z)t^a$. (The $x = (t,{\bf x})$ while $z$ is the fifth-dimension coordinate.)  The two representations produce seemingly different equations of motion and {\it potentially} different mass eigenvalues. Alternatively, one may use a linear representation, $X = X_{0} + i \pi^{a}t^a$, as in \cite{Kaplan:2009kr}. Any physical observables calculated from this model should of course be independent of the choice of representation.  

In this paper, we begin by recalling the modified gauge-gravity dual model in Section \ref{secDual}. For illustrative purposes, we examine the most prevalent exponential representation and then compare it to a linear representation. In Section \ref{secRep} we derive and explore the equations of motion of the two representations, showing that the sets of equations are, indeed, equivalent. One can simplify the equations of motion in two different ways: (i) by eliminating one field as in \cite{Sui:2009xe}, or (ii) by transforming the full system of second-order differential equations into Schr\"odinger-like form. Numerically calculating the mass eigenvalues in Section \ref{secPME}, we find that the eigenvalues found from one method do not match those of the other.  Apparently, \cite{Sui:2009xe} has found the eigenvalues not of the pseudoscalars but of a derivative field, and they are \textit{not} equivalent.  The pseudoscalar mass spectrum calculated from the model of \cite{Gherghetta:2009ac} agrees reasonably well with the observations.  In Section \ref{secGOR} we verify quantitatively that the pion mass-squared varies linearly with the quark mass, thus satisfying the Gell-Man--Oakes--Renner (GOR) relation.  We conclude with general comments on the pseudoscalar representation in Section \ref{secDiscuss}.

\section{The Dual Model} 
\label{secDual}

We investigate a modified version of a common soft-wall AdS/QCD model. The original model was introduced in \cite{Karch:2006pv} and further investigated in \cite{Evans:2006ea,Grigoryan:2008cc,Kwee:2007nq,Cherman:2008eh,Colangelo:2008us,Batell:2008zm}. A modified dilaton profile and quartic interactions were introduced in \cite{Gherghetta:2009ac} to separate explicit from spontaneous chiral symmetry breaking. The action for this AdS/QCD model is 
\begin{equation} \label{equAction1}
S_{5} = -\int{d^{5} x \sqrt{-g} {\rm e}^{-\chi(z)} Tr\left[|DX|^{2}+m_{X}^{2}|X|^{2} -\kappa |X|^{4}+\frac{1}{2 g_{5}^{2}}(F_{A}^{2}+F_{V}^{2})\right]}
\end{equation}
where the metric takes the form of 
\begin{equation}
ds^{2} = a(z)^{2}(\eta_{\mu\nu}dx^{\mu}dx^{\nu} + dz^{2})
\end{equation}
with the $(-++++)$ signature and the warp factor $a(z)=L/z$ with $z\geq0$. The covariant derivative is defined as $D_{M}=\partial_{M} + i [V_{M},X] - i\{A_{M},X\}$ (upper-case roman indices run from 0 to 4). The complex $2 \times 2$ field $X$ contains the scalar vacuum expectation value (VEV) $v(z)$, scalar field excitations $S(x,z)$, and the pseudoscalar field $\pi(x,z)$. Using the AdS/CFT dictionary, the scalar component is dual to the $\bar{q}q$ operator and, therefore, the scalar mass associated with this field in the 5-$D$ gravity theory is $m_{X}=-3/L^{2}$. We are only interested in a subset of terms related to the pseudoscalar sector,
\begin{eqnarray}
\mathcal{L} &=&  \sqrt{-g} {\rm e}^{-\chi(z)} Tr\left[-|DX|^{2}-m_{X}^{2}|X|^{2} + \kappa |X|^{4} -\frac{1}{2 g_{5}^{2}}(\partial_{M}A_{N} - \partial_{N}A_{M})^{2}\right] \nonumber \\
&=&  -\sqrt{-g} {\rm e}^{-\chi(z)} Tr\Big[g^{MN}(\partial_{M}X - i\{A_{M},X\} - i[V_{M},X])(\partial_{N}X^{\dagger} + i\{A_{N},X^{\dagger}\}+ i[V_{N},X^{\dagger}])\nonumber  \\
&&+m_{X}|X|^{2} -\kappa |X|^{4}+\frac{g^{MP}g^{NR}}{g_{5}^{2}}(\partial_{M} A_{N}\partial_{P} A_{R} - \partial_{M} A_{N}\partial_{R}A_{P})\Big] \label{equAction1simple}.
\end{eqnarray} 
A good phenomenological parameterization was described and justified in \cite{Gherghetta:2009ac},
\begin{eqnarray}
v(z) &=& \alpha z + \beta z \text{tanh}(\gamma z^{2}) \label{equprev}\\
\chi(z) &=& \int{dz\, \frac{\partial_{z}\left(a(z)^3 v'(z)\right)+a(z)^5 \left(\frac{1}{2} \kappa  v(z)^3+3 v(z)\right)}{a(z)^3 v'(z)}}, \label{equprechi}
\end{eqnarray}
where
\begin{eqnarray}
\alpha &=& \frac{\sqrt{3}m_{q}}{g_{5}L} \\
\beta &=& \sqrt{\frac{4\lambda}{\kappa L^{2}}} - \alpha\\
\gamma &=&  \frac{g_{5}\sigma}{\sqrt{3}\beta}.
\end{eqnarray}
For the rest of this paper we use the values quoted in \cite{Gherghetta:2009ac}: $m_q = 9.75$ MeV, $\sigma = (204.5$ MeV$)^{3}$, $\lambda = 0.18$ GeV$^2$ , $g_{5}=2\pi$, and $\kappa =15$.

Concentrating on the pseudoscalar sector, we explore two representations for the field $X$,
\begin{eqnarray}
X_e &=& (v(z)/2+S(x,z)) \, I \, {\rm e}^{2i \pi_{e}^{a}(x,z)t^{a}} \label{equXe}\\ 
X_l &=& (v(z)/2+S(x,z)) \, I + i \pi_{l}^{a}(x,z)t^{a} \label{equXl}
\end{eqnarray}
where $I$ is the $2\times 2$ identity matrix. We refer to $X_{e}$ as the exponential representation and $X_{l}$ as the linear representation. The exponential representation is used in \cite{Erlich:2005qh, Grigoryan:2008cc, Kwee:2007nq, DeFazio:2008mb}, where it is assumed to be the canonically normalized pion field, $\pi_{e} = \tilde{\pi}/f_{\pi}$. The representation $X_{e}$ is also used in \cite{Sui:2009xe}, where the eigenvalues of the pseudoscalar sector are computed using a method that we comment on later. The linear representation has been specified before in \cite{Kaplan:2009kr}, where the $\pi$ field carries the same dimensions as other fields in the Lagrangian. We already see an apparent difference between representations; (\ref{equXl}) allows for an explicit quartic term in $\pi$ when substituted into the action (\ref{equAction1}), whereas there is no such term in the case of (\ref{equXe}). The consequence of such quartic terms in $\pi$ will not be addressed here; however, the quartic term strength impacts the pseudoscalar mass spectrum through the parameter $\kappa$. We only consider field terms up to quadratic order.

\section{Representations} 
\label{secRep}

In this section we derive the equations of motion arising from the two representations (\ref{equXe}) and (\ref{equXl}). The pseudoscalar and longitudinal components of the axial-vector field mix in the Lagrangian; therefore, we find two coupled differential equations for each representation. This makes the numerical work more involved than for the scalar, vector, and axial-vector sections, which were already studied in \cite{Gherghetta:2009ac}.  The last part of this section shows that the two sets of differential equations are equivalent.

\subsection{Exponential Representation}
Let us take (\ref{equXe}) and substitute it into (\ref{equAction1simple}), where we focus only on the terms involving the field $\pi(x,z)$,
\begin{eqnarray}
\mathcal{L}_e &=&  -\sqrt{-g} {\rm e}^{-\chi(z)} \frac{1}{2}\delta^{ab}\Big(g^{MN} (v^{2}\,\partial_{M}\pi\partial_{N}\pi + v^{2} A_{M} A_{N}-2v^{2}\partial_{M}\pi A_{N})\nonumber\\
&&+ \frac{g^{MP}g^{NR}}{g_{5}^{2}}(\partial_{M} A_{N}\partial_{P} A_{R} - \partial_{M} A_{N}\partial_{R}A_{P}) \Big).\label{start}
\end{eqnarray}
We work in the axial gauge, $A_{z} = 0$, and define $A_{\mu} = A_{\mu\perp} + \partial_{\mu}\phi$, where $\partial_{\mu}A_{\perp}^{\mu}=0$. Separating (\ref{start}) explicitly into regular 4-$D$ components and extra-dimensional terms, we obtain
\begin{eqnarray}
\mathcal{L}_e &=& -\frac{1}{2} {\rm e}^{-\chi(z)}\Big[\sqrt{-g}g^{\mu\nu}(v^{2}\partial_{\mu}\pi\partial_{\nu}\pi+v^{2}A_{\mu}A_{\nu} - 2v^{2}\partial_{\mu}\pi A_{\nu})+\sqrt{-g}g^{zz}v^{2}\partial_{z}\pi\partial_{z}\pi\ \nonumber\\
&&+ \frac{\sqrt{-g}g^{\mu\nu}g^{\rho\sigma}}{g_{5}^{2}}(\partial_{\mu} A_{\rho} \partial_{\nu} A_{\sigma} - \partial_{\mu} A_{\rho} \partial_{\sigma} A_{\nu}) + \frac{\sqrt{-g}g^{zz}g^{\mu\nu}}{g_{5}^{2}}(\partial_{z} A_{\mu}\partial_{z} A_{\nu})\Big].\label{equLbase}
\end{eqnarray}
Keeping only terms of the longitudinal part of $A_{\mu}$ gives
\begin{eqnarray}
\mathcal{L}_{e} &=& -\frac{1}{2} {\rm e}^{-\chi(z)}\Big[\sqrt{-g}g^{\mu\nu}(v^{2}\partial_{\mu}\pi\partial_{\nu}\pi+v^{2}\partial_{\mu}\phi\partial_{\nu}\phi - 2v^{2}\partial_{\mu}\pi \partial_{\nu}\phi) \nonumber\\
&& + \sqrt{-g}g^{zz}v^{2}\partial_{z}\pi\partial_{z}\pi\ 
+ \frac{\sqrt{-g}g^{zz}g^{\mu\nu}}{g_{5}^{2}}(\partial_{z} \partial_{\mu}\phi\partial_{z} \partial_{\nu}\phi)\Big].\label{equLproper}
\end{eqnarray}
Varying (\ref{equLproper}) with respect to $\pi$ gives
\begin{equation}
\delta\mathcal{L}_e = \partial_{z} {\rm e}^{-\chi}\sqrt{-g}g^{zz}v^{2}\partial_{z}\pi \delta\pi + {\rm e}^{-\chi}\sqrt{-g}v^{2}g^{\mu\nu}\partial_{\nu}\partial_{\mu}(\pi - \phi)\delta\pi. \nonumber\\
\end{equation}
Using a Kaluza-Klein decomposition,
\begin{eqnarray}
\pi(x,z) &=& \sum_{n}\Pi_{n}(x) \pi_{n}(z) \label{equKKpi} \\
\phi(x,z) &=& \sum_{n}\Phi_{n}(x)\phi_{n}(z) \label{equKKphi}
\end{eqnarray}
and
\begin{equation}
\partial^{2} \Pi_{n}(x) = m_{n}^{2}\Pi_{n}(x) \, , \quad\quad \partial^{2}\Phi_{n}(x) = m_{n}^{2}\Phi_{n}(x) \,, 
\end{equation}
we can express the system of equations in terms of its $z$-dependent parts. We obtain the first equation of motion,
\begin{equation}
\frac{{\rm e}^{\chi}}{v^{2} a^{3}}\partial_{z}\left({\rm e}^{-\chi}v^{2}a^{3}\partial_{z}\pi_{n}\right) + m_{n}^{2}(\pi_{n} -\phi_{n}) = 0. \label{equOne}
\end{equation}
Varying (\ref{equLproper}) with respect to $\phi$ and breaking it down into KK modes gives the second equation of motion,
\begin{equation} \label{equTwo}
{\rm e}^{\chi}\partial_{z}\left(\frac{{\rm e}^{-\chi}}{z}\partial_{z}\phi_{n}\right) + \frac{g_{5}^{2}L^{2}v^{2}}{z^{3}}(\pi_{n}-\phi_{n}) = 0. 
\end{equation}

Alternatively, we can express (\ref{equOne}) and (\ref{equTwo}) in a Schr\"odinger-like form. We rewrite them in the same form as (\ref{equphiAppend}) and (\ref{equpiAppend}) in the Appendix by substituting
\begin{eqnarray}
\pi_{n} &=& {\rm e}^{f(z)}\tilde\pi_{n} \quad\quad\quad f(z) = \chi(z) + \log{\frac{z^{3}}{v(z)^{2}}} \\
\phi_{n} &=& {\rm e}^{g(z)}\tilde\phi_{n} \quad\quad\quad g(z) = \chi(z)+ \log{z},
\end{eqnarray}
which eliminate terms involving the first derivative of the fields, $\pi_{n}'$ and $\phi_{n}'$. After reverting back to the notation $\tilde\pi\rightarrow\pi$ and $\tilde\phi\rightarrow\phi$ the equations of motion become
\begin{eqnarray}
&& -\pi_{n}'' + \left(\frac{\chi'^{2}}{4} - \frac{\chi''}{2}-\frac{\chi'v'}{v} + \frac{3\chi'}{2z} + \frac{15}{4z^{2}}-\frac{3 v'}{v z}+\frac{v''}{v}-m_{n}^{2}\right)\pi_{n} = -m_{n}^{2}\frac{v^{2}L^{2}}{z^{2}}\phi_{n} \label{equSchexppi}\\
&& -\phi_{n}'' + \left(\frac{\chi'^{2}}{4} - \frac{\chi''}{2}+\frac{\chi'}{2z} + \frac{3}{4 z^{2}} + \frac{g_{5}^{2}v^{2}L^{2}}{z^{2}}\right)\phi_{n} = g_{5}^{2} \pi_{n} \label{equSchexpphi}
\end{eqnarray}
where ($'$) indicates the derivative with respect to $z$.

\subsection{Linear Representation}
 
When considering the linear representation of the pseudoscalar field (\ref{equXl}), we find quadratic and quartic $\pi$ terms that were not explicitly present in the exponential representation. After making the appropriate substitutions, we find that
\begin{eqnarray}
\mathcal{L}_l &=& -\frac{1}{2} {\rm e}^{-\chi}\sqrt{-g}\Big(g^{\mu\nu}\partial_{\mu}\pi\partial_{\nu}\pi + g^{zz}\partial_{z}\pi\partial_{z}\pi - 2 v g^{\mu\nu}\partial_{\mu}\pi\partial_{\nu}\phi + m_{X}^{2}\pi^{2} -\frac{\kappa}{2}v^{2}\pi^{2} \nonumber\\
&+& g^{\mu\nu}v^{2} \partial_{\mu}\phi\partial_{\nu}\phi + \frac{g^{\mu\nu} g^{zz}}{g_{5}^{2}}\partial_{z}\partial_{\mu}\phi\partial_{z}\partial_{\nu}\phi\Big).
\end{eqnarray} 
Once again, we derive two coupled equations. Varying with respect to $\phi$ produces a result similar to $X_{e}$ with the exception of factors of the VEV in the mixing term, giving
\begin{equation}\label{equphi}
{\rm e}^{\chi}\partial_{z}\left(\frac{{\rm e}^{-\chi}}{z}\partial_{z}\phi_{n}\right) + \frac{g_{5}^{2}L^{2} v}{z^{3}}\left(\pi_{n} - v \phi_{n}\right) = 0.
\end{equation}
Varying with respect to $\pi$ gives the second equation of the linear representation,
\begin{equation}\label{equpi}
z^{3}{\rm e}^{\chi}\partial_{z}\left(\frac{{\rm e}^{-\chi}}{z^{3}}\partial_{z}\pi_{n}\right) - \left(\frac{m_{X}^{2}}{z^{2}} - \frac{\kappa L^{2} v^{2}}{2 z^{2}}\right)\pi_{n}+m_{n}^{2}\pi_{n} = m_{n}^{2}v\phi_{n}.
\end{equation}
We can express (\ref{equphi}) and (\ref{equpi}) in a Schr\"odinger-like form with the substitutions,
\begin{eqnarray}
\pi_{n} &=& {\rm e}^{f}\tilde{\pi_{n}}\quad\quad\quad f = \frac{\chi}{2} + \frac{3}{2} \log{\frac{z}{L}} \\
\phi_{n} &=& {\rm e}^{g}\tilde{\phi_{n}}\quad\quad\quad g = \frac{\chi}{2} + \frac{1}{2} \log{\frac{z}{L}}.
\end{eqnarray}
Simplifying the equations and reverting back to the notation $\tilde{\pi_{n}}\rightarrow\pi_{n}$ and $\tilde{\phi_{n}}\rightarrow\phi_{n}$ for simplicity, we find
\begin{eqnarray}
&&-\phi_{n}'' + \left(\frac{\chi'^{2}}{4} - \frac{\chi''}{2} + \frac{3}{4z^{2}} + \frac{\chi'}{2 z} + \frac{g_{5}^{2}L^{2}v^{2}}{z^{2}}\right)\phi_{n} = \frac{g_{5}^{2}L v}{z}\pi_{n}\label{equSchphi} \\
&&-\pi_{n}'' + \left(\frac{\chi'^{2}}{4} - \frac{\chi''}{2} + \frac{3}{4z^{2}} + \frac{3\chi'}{2 z} - \frac{\kappa L^{2}v^{2}}{2z^{2}} - m_{n}^{2} \right)\pi_{n} = -m_{n}^{2}\frac{vL}{z}\phi_{n}  \label{equSchpi}
\end{eqnarray}

\subsection{Representation Equivalence}
 
The pseudoscalar field representation should not affect the physical results obtained from the model. Examining the two sets of coupled equations in each representation, we see that neither (\ref{equOne}) nor (\ref{equTwo}) contains an explicit dependence on $\kappa$, whereas (\ref{equpi}) does. Although dependence on $\kappa$ does not appear explicitly in the exponential representation, it does enter through the function $v(z)$.

We begin by expanding $X_{e}$,
\begin{eqnarray}
X_{e} &=& \left(\frac{v}{2} + S\right)(1 + 2 i \pi_{e} + \ldots)\nonumber \\
&=& \frac{v}{2} + S + i\pi_{e} v. \label{equXexpand}
\end{eqnarray}
Comparing (\ref{equXexpand}) to (\ref{equXl}), we surmise that $\pi_{e} v(z)\rightarrow \pi_{l}$ relates the two representations. Let us substitute $\pi_{e}\rightarrow \pi_{l}/v(z)$ into the equations of motion of the exponential representation and attempt to obtain the equations of motion of the linear representation. The substitution into (\ref{equTwo}) is trivial; it yields
\begin{equation}
{\rm e}^{\chi}\partial_{z}\left(\frac{{\rm e}^{-\chi}}{z}\partial_{z}\phi\right) + \frac{g_{5}^{2}v}{z^{3}}(\pi_{l}- v\phi) = 0,
\end{equation}
which is equivalent to (\ref{equphi}) as expected. Showing the equivalence of the other two equations requires a bit more work. First, we substitute for $\pi_{e}$ in (\ref{equOne}) and then simplify the expression,
\begin{equation} 
\frac{z^{3}{\rm e}^{\chi}}{v}\partial_{z}\left(\frac{{\rm e}^{-\chi}v^{2}}{z^{3}}\left(\frac{\pi_{l}'}{v} - \frac{\pi_{l} v}{v^{2}}\right)\right) + m_{n}^{2}(\pi_l -v\phi) = 0
\end{equation}
which becomes
\begin{equation}
\pi_{l}'' - \left(\chi' + \frac{3}{z}\right)\pi_{l}' - \frac{\pi_l}{v}\left(v'' - \chi'v' - \frac{3}{z}v'\right) + m_{n}^{2}(\pi_l -v\phi) = 0. \label{equOnemid}
\end{equation}
Recall the equation of motion for $v(z)$, found in \cite{Gherghetta:2009ac}, which can be derived from (\ref{equAction1}) and does not depend on the representation.
\begin{equation} \label{equv}
v'' - \left(\chi'+\frac{3}{z}\right)v' + \left(\frac{3}{z^{2}} + \frac{\kappa L^{2}v^{2}}{2 z^{2}}\right)v = 0.
\end{equation}
Using (\ref{equv}) in (\ref{equOnemid}), we find
\begin{equation}
\pi_{l}'' - \left(\frac{3}{z}+\chi'\right)\pi_{l}' + \left(\frac{3}{z^{2}} + \frac{\kappa L^{2}v^{2}}{2 z^{2}}\right)\pi_{l} + m_{n}^{2}\left(\pi_{l} -  v\phi\right) = 0,
\end{equation}
which is the same as the equation of motion of the linear representation (\ref{equpi}). In a similar way, this equivalence can be shown by starting with the linear representation and substituting $\pi_{l} = v(z)\pi_{e}$.

\section{Pseudoscalar Mass Eigenvalues}
\label{secPME}

We investigate two ways to calculate the pseudoscalar eigenvalues $m_{n}^{2}$. Rearranging and eliminating the longitudinal component $\phi$ is one strategy outlined in \cite{Sui:2009xe} and is briefly presented here. Alternatively, the Appendix contains a numerical routine we use to solve the set of coupled equations. Using this method, we find that $\pi_{e}$, the ratio of $\pi_{l}$ and $v(z)$, is extremely sensitive to boundary conditions, the reason being that $v(z)$ goes to zero as $z$ goes to zero.  This makes resolution of the eigenvalues difficult and subject to significant numerical error. Fortunately, we have shown explicitly that physical results do not depend on the particular representation of the pseudoscalars. Therefore we determine the eigenvalues in the linear representation. 

First, we follow the method of \cite{Sui:2009xe}. We manipulate (\ref{equOne}) and (\ref{equTwo}) to eliminate the $\phi$ field. Adding (\ref{equOne}) and (\ref{equTwo}) yields
\begin{equation}\label{equOneMod}
g_{5}^{2}a^{2}v^{2}\partial_{z}\pi_{n} = m_{n}^{2}\partial_{z}\phi_{n} \, .
\end{equation}
We then use (\ref{equOneMod}) to replace the first term in (\ref{equTwo}) and solve for $\phi$,
\begin{eqnarray} 
\phi_{n} &=& \frac{1}{g_{5}^{2}h(z)}\partial_{z}\left[a {\rm e}^{-\chi}\left(-\frac{g_{5}^{2}}{q^{2}}a^{2}v^{2}\partial_{z}\pi_{n}\right)\right] + \pi_{n}\nonumber \\
m_{n}^{2}\partial_{z}\phi_{n} &=& \partial_{z}\left(h(z)^{-1}\partial_{z}(h(z)\partial_{z}\pi_{n})\right) + m_{n}^{2}\partial_{z}\pi_{n} 
\end{eqnarray}
where $h(z) = a(z)^{3}v^{2} {\rm e}^{-\chi}$. Using (\ref{equOneMod}) again, we denote $\partial_{z}\pi\rightarrow \tilde\pi$ and rearrange to find the eigenvalue equation
\begin{equation}\label{equTildepi}
-\partial_{z}[h^{-1}\partial_{z}(h \tilde\pi_{n})] + g_{5}^{2}a ^{2}v^{2}\tilde\pi_{n} = m_{n}^{2} \tilde\pi_{n} \,.
\end{equation}
This can be put into the Schr\"odinger-like form by substituting $\Pi = \tilde\pi/\sqrt{h(z)}$.  Then (\ref{equTildepi}) becomes
\begin{equation}\label{equMassEigen}
-\Pi_{n}'' + V(z)\Pi_{n} = m_{n}^{2} \Pi_{n}
\end{equation}
where the potential takes the form,
\begin{eqnarray}
V(z) &=& \frac{3h'^{2}}{h^{2}}-\frac{h''}{2h}+\frac{g_{5}^{2}L^{2}v^{2}}{z^{2}}\nonumber \\
&=& \frac{3}{4 z^{2}} -\frac{3 v'}{z v} + 2\frac{v'^{2}}{v^{2}} + \frac{3\chi'}{2 z} - \frac{v'\chi'}{v} + \frac{\chi'^{2}}{4} - \frac{v''}{v} + \frac{\chi''}{2} + \frac{g_{5}^{2}L^{2}v^{2}}{z^{2}} \,. \label{equPot}
\end{eqnarray}
Solving (\ref{equMassEigen}) using a standard shooting method gives the mass spectrum shown in Table \ref{tblmass}. There is no low-mass Goldstone boson and no large mass gap between the first two eigenvalues. What we have done is taken 2 second-order differential equations and reduced them to 1 third-order differential equation and found the eigenvalues of $\partial_{z}\pi$. We seem to lose the Goldstone boson using this method. 

\begin{figure}
\begin{center}
\includegraphics[scale=0.45]{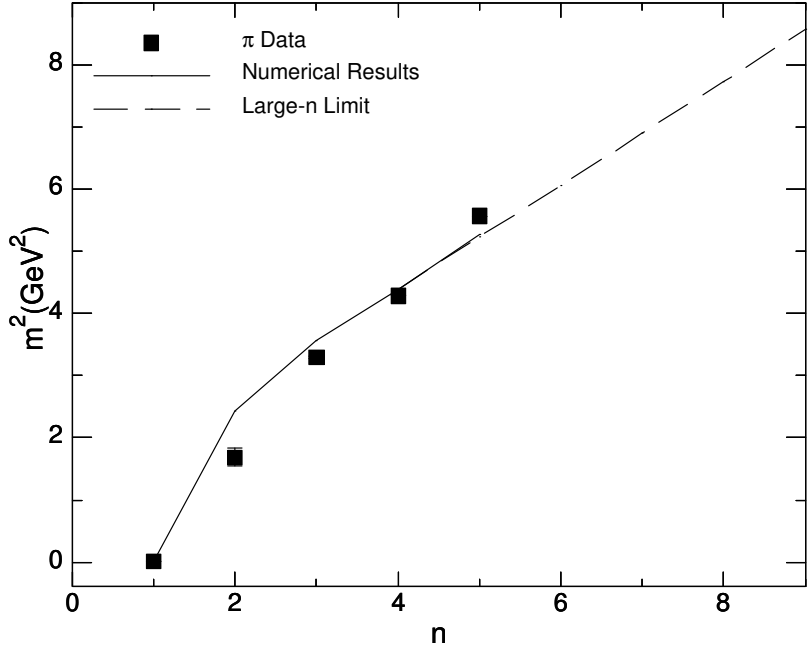}
\caption{The mass spectrum calculated in the AdS/QCD model is plotted along with the experimental data \cite{2008pdg}. The eigenvalues display two characteristics matching the QCD pion spectrum: (1) low-mass ground state and (2) a large gap between the ground state and the first excited state. The large-$n$ mass trajectory clearly follows our calculated eigenvalues from $n\approx 4$ when our numerical routine inadequately follows the oscillations of the higher eigenfunctions.}
\label{linEigen}
\end{center}
\end{figure}

The numerical routine described in the Appendix calculates the mass eigenvalues of (\ref{equSchphi}) and (\ref{equSchpi}), which are then plotted in Figure \ref{linEigen} and listed in Table \ref{tblmass}. Solving the set of equations directly produces a mass spectrum with a Goldstone boson and with mass eigenstates that match well with the observed radial pion excitations. These results show that eliminating one of the fields from the system of second-order differential equations also eliminates information from the mass spectrum. Thus, the eigenvalues of $\partial_{z}\pi_{e}$ do not match those of $\pi_{l}$ or $\pi_{e}$, at least not in this model. The validity of equating the two sets of eigenvalues in \cite{Sui:2009xe} may need to be reassessed.

\begin{table}
\begin{center}
\begin{tabular}{| l | c | c | c | c |}
\hline
n & $\pi$ Data (MeV) 	& $\pi_{l}$ (MeV) 		& Large-$n$ $\pi_{l}$ 	& $\partial_{z}\pi_{e}$ (MeV) \\
\hline
1 & 140  		& 143  				& -			& 1440\\
2 & 1300 $\pm$ 100 	& 1557  			& -			& 1706 \\
3 & 1816 $\pm$ 14 	& 1887  			& -			& 1925\\
4 & 2070* 		& 2095 				& - 			& 2117\\
5 & 2360* 		& 2298 				& 2245 			& 2290\\
6 &  -   		& - 				& 2403 			& 2451\\
7 &  -   		& -				& 2551 			& 2601\\
\hline
\end{tabular}
\caption{The observed masses \cite{2008pdg} and calculated masses using the linear representations. The large-$n$ limit solutions are valid from $n\approx 4$. From that point onward, the numerical method used becomes increasingly inaccurate and tends to skew the $\pi_{l}$ eigenvalues to larger values than are expected from the linear behavior. The eigenvalues found using the method of \cite{Sui:2009xe} are also shown.
 *Appears strictly in the further states of \cite{2008pdg}.}
\label{tblmass}
\end{center}
\end{table}

For large-$n$ excitations, the technique described in the Appendix runs into boundary condition problems for $n\geq 4$. As the number of oscillations in the eigenfunctions increases for higher $n$ modes, the routine finds eigenvalues that are skewed to larger values. To uncover the correct asymptotic behavior for large $n$, we take the large-$z$ limit of (\ref{equSchphi}) and (\ref{equSchpi}). As $n$ increases, the eigenfunction is largely determined by the behavior of the effective potential at large $z$. At large $z$, the dilaton and tachyon behave as
\begin{eqnarray}
v(z) &=&  (\alpha+\beta)z \equiv \Gamma \frac{z}{L}\\
\chi(z) &=& \lambda z^{2}.
\end{eqnarray}
To take the large-$z$ limit of both representations, we introduce a new dimensionless parameter, $\xi=\sqrt{\lambda}z$, and expand in $\xi$. In the exponential representation, we find that (\ref{equSchexppi}) and (\ref{equSchexpphi}) become
\begin{eqnarray}
-\pi_{k}'' + \xi^{2}\pi_{k}  &=& \frac{m_{k}^{2}}{\lambda}\left(\pi_{k} - \Gamma\phi_{k}\right) \label{equpiExpSHO}\\
-\phi_{k}'' + \xi^{2}\phi_{k} &=& \frac{g_{5}^{2}}{\lambda}\left(\pi_{k} - \Gamma\phi_{k}\right). \label{equphiExpSHO}
\end{eqnarray}
Similarly, in the linear representation the expansion of (\ref{equSchphi}) and (\ref{equSchpi}) at large $\xi$ yields the eigenvalue equations
\begin{eqnarray}
-\pi_{k}'' + \xi^{2} \pi_{k}&=& \left(\frac{\kappa\Gamma^{2}}{2 \lambda} -2 + \frac{m_{k}^{2}}{\lambda}\right)\pi_{k} - \frac{m_{k}^{2}\Gamma}{\lambda}\phi_{k} \label{equpiSHO} \\
-\phi_{k}'' + \xi^{2} \phi_{k} &=& \frac{g_{5}^{2}\Gamma}{\lambda}\left(\pi_{k} - \Gamma\phi_{k}\right),\label{equphiSHO}
\end{eqnarray}
where ($'$) indicates differentiation with respect to $\xi$. Each set of equations appears to describe a pair of simple harmonic oscillators, the equations of motion of which are
\begin{eqnarray}
-\phi_{k}'' + \xi^{2} \phi_{k} &=& (2k+1)\phi_{k} \label{equphiSHOG} \\
-\pi_{k}'' + \xi^{2}\pi_{k} &=& (2k+1)\pi_{k}\quad\quad k=0,1,\ldots. \label{equpiSHOG}
\end{eqnarray}
It is a reasonable assumption that $\phi_{k} = c_{k}\pi_{k}$; it ensures that (\ref{equpiExpSHO}), (\ref{equphiExpSHO}), (\ref{equpiSHO}), and (\ref{equphiSHO}) have solutions. Using the form of (\ref{equphiSHOG}) and (\ref{equpiSHOG}) to solve for $c_k$ and $m_{k}^{2}$ in both representations, we find
\begin{eqnarray}
c_{k} &=& \frac{g_{5}^{2}}{g_{5}^{2}\Gamma^{2} +(2k+1)\lambda} \label{equexpcn}\\
m_{k}^{2} &=& (2k+1)\lambda + g_{5}^{2}\Gamma^{2}
\end{eqnarray}
for the exponential representation and 
\begin{eqnarray}
c_{k} &=& \frac{g_{5}^{2}\Gamma}{g_{5}^{2}\Gamma^{2} +(2k+1)\lambda} \label{equlinearcn}\\
m_{k}^{2} &=& \frac{\left((2k+3)\lambda -\frac{1}{2}\kappa\Gamma^{2}\right)\left(g_{5}^{2}\Gamma^{2} + (2k+1)\lambda\right)}{(2k+1)\lambda} \label{equlinearmn}
\end{eqnarray}
for the linear representation. 

So far, we have neglected an important fact:  $z\geq 0$. The eigenfunctions $\phi$ and $\pi$ describe \textit{half}-harmonic oscillators and contain only half the modes that full harmonic oscillators do; therefore, we must take $k\rightarrow 2k$. From \cite{Gherghetta:2009ac} we have 
\begin{equation}
\Gamma^{2} = \frac{4 \lambda}{\kappa}.
\end{equation} 
The mass eigenvalues for large $n$, where $n=k+1$, in both representations then become 
\begin{equation} \label{equMass4Both}
m_{n}^{2} = (4 n-3)\lambda + g_{5}^{2}\Gamma^{2}\quad\quad n=4,5,\ldots
\end{equation}
which are also listed in Table \ref{tblmass} and plotted in Figure \ref{linEigen}. Combining (\ref{equMass4Both}) and the numerical technique, we obtain all the pseudoscalar eigenvalues. By simple investigation, we find that the large-$n$ eigenvalues should be trusted over the ones found with the numerical routine for $n \geq 4$.

\section{Gell-Mann--Oakes--Renner Relation} 
\label{secGOR}

In this section we explore the Gell-Mann--Oakes--Renner relation numerically.  Using the established equivalence between the exponential and linear representations, $\pi_e = \pi_l/v(z)$, and inserting it into (\ref{equOneMod}), we obtain
\begin{equation}
\frac{g_5^2L^2v^2}{z^2} \partial_z\left(\frac{\pi_l}{v}\right) = m_\pi^2\partial_z\phi \,.
\end{equation}
Following the method of \cite{Erlich:2005qh}, we construct a perturbative solution in $m_{\pi}$ where $\phi(z) = A(0,z)-1$ and use 
\begin{equation}
f_\pi^2=\left. -L \frac{\partial_z A(0,z)}{g_5^2z}\right |_{z\rightarrow0}. \label{eq:fpi}
\end{equation}
From this, it follows that
\begin{equation}
\pi(z)=m_\pi^2\,v(z)\int_0^z du\, \frac{u^3}{v^2(u)} \frac{\partial_z A(0,u)}{g_5^2u} \,.
\end{equation}
The function $u^3/v^2(u)$ is significant only at small values of $u\sim\sqrt{m_q/\sigma}$, where we may use (\ref{eq:fpi}) to relate the derivative on $A(0,u)$ to the pion decay constant, so that
\begin{equation} \label{equPreGOR}
\frac{\pi_l}{v}=-\frac{m_\pi^2 f_\pi^2}{2m_q\sigma} \,.
\end{equation}
We find that $\pi_l=-v(z)$ solves the axial-vector field's equation of motion found in \cite{Gherghetta:2009ac}
\begin{equation}
{\rm e}^\chi\partial_z\left(\frac{{\rm e}^{-\chi}}{z}\partial_z A_\mu(q,z)\right)-\frac{q^2}{z}A_\mu(q,z) - \frac{g_5^2L^2v^2}{z^3} A_\mu(q,z) = 0
\end{equation}
in the region of small $z$ and as $q\rightarrow 0$. As a result, (\ref{equPreGOR}) becomes the expected Gell-Mann--Oakes--Renner (GOR) relation,
\begin{equation}
2m_q\sigma=m_\pi^2 f_\pi^2 \,.
\end{equation}
This perturbative behavior for $\pi_{e}$ and $\pi_l$ justifies the use of Neumann and Dirichlet boundary conditions, respectively, such that
\begin{equation}
\pi_{e}(0) = -1 \,, \quad\quad\quad \pi_{l}(0) = v(0) = 0 \,.
\end{equation}

\begin{figure}
\begin{center}
\includegraphics[scale=0.45]{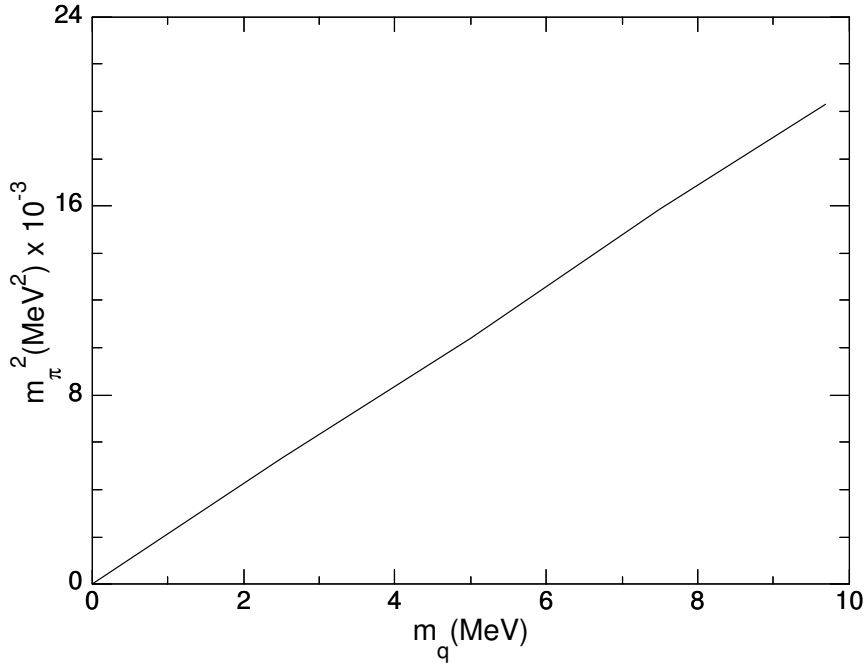}
\caption{Plot of $m_{\pi}^{2}$ versus $m_{q}$ produces a straight line from which the pion decay constant $f_{\pi}$ is calculated using (\ref{equGORform2}).}
\label{figGOR}

\end{center}
\end{figure}

The ratio of $m_{q}$ and $m_{\pi}^{2}$ should be a constant following the GOR relation.
\begin{equation} \label{equGORform2}
\frac{m_{q}}{m_{\pi}^{2}} = \frac{f_{\pi}^{2}}{2\sigma} 
\end{equation}
We solve the pair of coupled differential equations for the ground-state pseudoscalar mass, $m_{\pi}$, for differing values of $m_{q}$ to ensure that the numerical routine of the Appendix respects the GOR relation. The results are plotted in Figure \ref{figGOR}. We see linear behavior in the plot, indicating that as $m_{q}\rightarrow 0$ we obtain a constant ratio of $m_{q}/m_{\pi}^{2}$. The slope of the line in Figure \ref{figGOR} implies $f_{\pi} \approx 90$ MeV, a result consistent with the input parameters as described in \cite{Gherghetta:2009ac}. 

\section{Discussion} 
\label{secDiscuss}

We investigated two possible representations of the pseudoscalar field within a model of AdS/QCD that incorporates both explicit and spontaneous chiral symmetry breaking as well as confinement.  We showed formally the equivalence of these representations. The Gell-Mann--Oakes--Renner relation was shown to hold both analytically and numerically.  We also found that the eigenvalues of the derivative field $\partial_{z}\pi_{e}$ are not the same as those of the field $\pi_{e}$, as \cite{Sui:2009xe} suggests. 

This paper completes the calculation of the radial mass spectra of the spin-0 and spin-1 fields in the soft-wall model of \cite{Gherghetta:2009ac}. The pseudoscalar fits into the AdS/QCD model quite well, reproducing a number of pion features, namely, (1) a low-energy ground state corresponding to the Goldstone boson, (2) a relatively large gap between the ground state and the first excited state, (3) a linear trajectory, and (4) a constant ratio of $m_{q}/m_{\pi}^{2}$ that gives a value of $f_{\pi}$ within a few percent of its accepted value.

\paragraph{Acknowledgments}
TK thanks Brian Batell and Todd Springer for insightful discussions. TK also thanks Daniel Sword for help with the numerical techniques used in this paper. This work was supported by the US Department of Energy (DOE) under Grant No. 
DE-FG02-87ER40328.

\begin{appendix}
\section{Numerical Routine} \label{appRoutine}

The equations of motion can be reduced to a set of second-order differential equations
\begin{eqnarray}
-\phi'' + V_1(z)  \phi + f(z)\pi  = 0 \label{equphiAppend}\\
-\pi'' + V_2(z) \pi + g(z)\phi = 0 \label{equpiAppend}
\end{eqnarray}
where the eigenvalues are contained within the coefficient functions. These equations can be reexpressed as a system of first-order differential equations
\begin{equation} \label{equmatrixphi}
\Phi' + W(z) \Phi = 0
\end{equation}
where $W$ is the matrix
\begin{equation}
W = \left(\begin{array}{cccc}
0     & 1     & 0     & 0 \\ 
V_1(z)& 0     &   f(z)& 0 \\
0     & 0     & 0     & 1 \\
g(z)  & 0     & V_2(z)& 0
\end{array}\right)
\end{equation}
and $\Phi$ is the vector
\begin{equation}
\Phi_{\alpha i} = \left(\begin{array}{c} \phi_{i} \\
			-\phi'_{i} \\
			\pi_{i} \\
			-\pi'_{i} \end{array}\right)		
\end{equation}
that forms an orthonormal basis of solutions. We can propagate the solution $\Phi$ between two boundary points
\begin{equation}
\Phi(z_1) = U(z, z_1, z_0, m_n^2)\Phi(z_0)
\end{equation}
where we solve (\ref{equmatrixphi}) with the appropriate boundary condition at $z_0$. The eigenvectors and eigenvalues of $U$ are then calculated. We find two large and two small eigenvalues corresponding to two nonrenormalizable and two normalizable eigenfunctions, respectively. Let us assume the eigenvectors $u_3$ and $u_4$ correspond to the small eigenvalues, $\lambda_3$ and $\lambda_4$. Then, any solution for $\Phi_{i}$ can be written as 
\begin{equation}
\Phi_{i} = \alpha u_3 + \beta u_4,
\end{equation}   
where we take the boundary condition as 
\begin{equation}
\Phi_{i}(z_0) =\left( \begin{array}{c} \phi(z_0) \\ -\phi'(z_0) \\ \pi(z_0) \\ -\pi'(z_0) \end{array}\right) \, . 
\end{equation}
In order for $\alpha$ and $\beta$ to be nontrivial, we must satisfy 
\begin{equation}\label{equNontrivial1}
\left(\begin{array}{cc} u_3^1 & u_4^1 \\ u_3^{3} & u_4^{3} \end{array}\right)\left(\begin{array}{c} \alpha \\ \beta\end{array}\right) = 0
\end{equation}
for Dirichlet or
\begin{equation}\label{equNontrivial2}
\left(\begin{array}{cc} u_3^2 & u_4^2 \\ u_3^{4} & u_4^{4} \end{array}\right)\left(\begin{array}{c} \alpha \\ \beta\end{array}\right) = 0
\end{equation}
for Neumann boundary conditions. We do this by cycling through eigenvalues $m_n^2$ that minimize the determinant of the 2$\times$2 matrix in (\ref{equNontrivial1}) or (\ref{equNontrivial2}). Practically, we find the singular points in the graph of the quantity $u_3^1 u_4^3 - u_3^3 u_4^1$ (or $u_3^2 u_4^4 - u_3^4 u_4^2$) versus $m_n^2$. An abrupt change in its behavior signals an eigenvalue. Of course, the elements chosen from the eigenvectors $u_3$ and $u_4$ are dependent upon the choice of Neumann or Dirichlet conditions on the boundary $z_0$.
\end{appendix}

\bibliographystyle{h-physrev4}
\bibliography{piRef}

\end{document}